\documentclass[12pt]{revtex4}

\usepackage{amsfonts,wrapfig}
\usepackage{amssymb}

\usepackage{amsfonts}
\usepackage{amsmath}
\usepackage{amssymb}
\usepackage{amsthm}
\usepackage{epsf}
\usepackage[dvips]{graphicx}
\usepackage{graphicx}

\def\bk{{\mbox{\boldmath$k$}}}

\def\bp{{\mbox{\boldmath$p$}}}
\def\bgam{{\mbox{\boldmath$\gamma$}}}

\begin{document}

\begin{center}
 A NOVEL APPROACH IN SOLVING  THE  SPINOR-SPINOR BETHE-SALPETER  EQUATION
 \end{center}

 \centerline{S.M. Dorkin$^{a,c}$ ,
 M. Beyer$^b$, S.S. Semikh$^c$  and  L.P. Kaptari$^c$}

$^a${International University Dubna, Dubna, Russia

 $^b$Institute
of Physics University of Rostock, Rostock, Germany

 $^c$ Bogoliubov
Lab. Theor. Phys. Joint Institute for Nuclear Research, Dubna,
Russia

\begin{abstract}
To solve the spinor-spinor  Bethe-Salpeter equation in   Euclidean
space we propose a novel method related to the use of
hyperspherical harmonics.  We suggest an appropriate extension to
form a new basis of spin-angular harmonics that is suitable for a
representation of the  vertex functions. We present a numerical
algorithm to  solve  the Bethe-Salpeter equation and investigate
in detail the properties of the solution for the scalar,
pseudoscalar and vector  meson exchange kernels including the
stability of bound states. We also compare our results to the non
relativistic ones and to the results given by light front
dynamics.
\end{abstract}

\section{Introduction}
Interpretation of many modern experiments
requires a covariant description of the two-body system. This is
either due to high precision that calls for an inclusion of all possible
corrections to a standard (possibly nonrelativistic) approach or due
to the high energies and momenta involved in the processes investigated.

In the spirit of a local quantum field theory the starting point of a relativistic
covariant description of  bound states of two particles is the
Bethe-Salpeter (BS)  equation. However, despite the obvious simplicity of
 two-body systems, the procedure of solving the BS equation
 encounters difficulties. These are related to singularities
 and branch points (cuts) of the  amplitude
 along the  real axis of the relative
 energy in  Minkowski space. Therefore, up to now the BS equation
including realistic interaction kernels has been solved either in Euclidean space
 within the ladder approximation or utilizing additional approximations
 of the equation itself.

Unfortunately, our understanding of the mathematical properties of
bound states within a relativistic approach is far from being
perfect. In mathematical terms the BS equation itself is a quite
complicated object, and the technical problem of solving it is
still a fundamental issue. Consequently there are very few
successful examples of solving the BS equation for fermions
including realistic interactions. For example, in
Ref.~\cite{umnikov} the BS equation for spinor particles was
regularly treated by using a two-dimensional Gaussian mesh. That
series of studies revealed a high feasibility of the BS approach
to describe nucleon-nucleon interactions, in particular, processes
involving the  deuteron. However, it should be mentioned that the
algorithms exploiting the two-dimensional meshes are rather
cumbersome and require large computer resources. In addition, the
numerical solution is obtained as two-dimensional arrays  which
are quite awkward in practice when computing matrix elements and
in attempts to establish  reliable parameterizations and possible
analytical continuations of the solution back to  Minkowski space.
 Therefore, it
is necessary to provide  a method for solving BS equations
that would feature a smaller degree of arbitrariness.

In the present paper we suggest an efficient  method to solve the
BS equation for fermions involving interaction kernels of
one-boson exchange type supplemented by corresponding form
factors. It is based on hyperspherical harmonics used to expand
partial amplitudes and kernels. We show that this novel technique
provides many insights into the BS approach. The current study is
partially stimulated by the results reported in
Ref.~\cite{karmanov}. We explore the structure of $^1S_0$ and
$^3S_1 - ^3D_1$ bound states for different couplings and study the
details of the convergence of solutions and corresponding
eigenvalues. In particular, on the basis of the proposed   method
for solving the BS equation it becomes possible to analyze the
specifics of the problem related to the stability of bound states in the BS approach.
Besides, the hyperspherical expansion provides an effective
parameterization of the  amplitude, which is extremely useful in
practical calculations of observables and in theoretical
investigations  of  the separability of the BS kernel with
one-boson exchange interaction. The detailed description of method
may be found in~\cite{our-FB}.

\section{Overview of the method}

 The BS equation for two spinor particles interacting via
one-boson-exchange potentials for
  vertex ${\cal G}(p)$ being  a $4\times 4$ matrix in
spinor space has the form
\begin{eqnarray}
\label{sphom} {\cal G}(p)=i g^2
\int\frac{d^4k}{(2\pi)^4}\,V(p,k)\,\Gamma(1)\, S(k_1)\,{\cal
G}(k)\,{\tilde S(k_2)}\,\tilde\Gamma(2),
\end{eqnarray}
 where  the propagator $V(p,k)$
 for  scalar and pseudoscalar exchange mesons is
\begin{eqnarray}
V(p,k)=\frac{1}{(p-k)^2-\mu^2+i\varepsilon},\label{vecprop}
\end{eqnarray}
while for  spinor particles the corresponding propagators
  are
\begin{eqnarray}
\nonumber {S}(k_1)= \frac{\hat k_1
+m}{k_1^2-m^2+i\varepsilon},\quad {\tilde S}(k_2)\equiv
C{S}(k_2)^TC=\frac{\hat k_2 -m}{k_2^2-m^2+i\varepsilon},
\end{eqnarray}
 where $C=i\gamma^0\gamma^2$ is  the charge conjugation matrix, $k_{1,2}$ and $p_{1,2}$
 are the four-momenta of the constituent particles, $k$ and $p$ are the corresponding
  relative momenta. The meson vertices $\Gamma(1,2)$ are determined by the
effective interaction Lagrangians urged to describe the considered fermion system.
 For a  two nucleon system, within the one-boson exchange approximation,
 these  vertices   are
$\Gamma(1)= 1;$ $\tilde\Gamma(2)=- 1$ for scalar and
$\Gamma(1)=\gamma_5;\,\, \tilde\Gamma(2)=-\gamma_5$
for pseudoscalar couplings, respectively.
 Each interaction
vertex $\Gamma$ is augmented with  cut-off monopole form factors
$F(q^2)= {\Lambda^2}/{(\Lambda^2-q^2)}$
 where $\Lambda$ are free  parameters. Note that the coupling constant $g$ in eq.~(\ref{sphom}) is
purely imaginary  for the pseudoscalar mesons else purely real.

The first step is the  expansion  of the vertex function ${\cal G}(p_0,\bp)$
into spin-angular harmonics
\begin{eqnarray}
{\cal G}(p_0,\bp)&=&\sum\limits_\alpha g_\alpha(p_0,|\bp|) \,{\cal
T}_\alpha(\bp)\label{spex}.
\end{eqnarray}
For specific bound states with given quantum numbers only some
basis matrices contribute to the vertex function  ${\cal G}(p)$ .
E.g., for the $^1S_0$ state only four matrices are relevant to
describe the amplitude, while in the $^3S_1-^3D_1$ channel eight
basis matrices are needed.
 In the $^1S_0$ channel
the  basis is
\begin{eqnarray}
{\cal T}_1(\bp)=\frac{1}{\sqrt{16\pi}}\gamma_5;\
{\cal T}_2(\bp)=\frac{1}{\sqrt{16\pi}}\gamma_0\gamma_5;\
{\cal T}_3(\bp)=\frac{1}{\sqrt{16\pi}}\frac{(\bp \bgam)}{|\bp|}\gamma_0\gamma_5;\
{\cal
T}_4(\bp)=\frac{1}{\sqrt{16\pi}}\frac{(\bp\bgam)}{|\bp|}\gamma_5.
\label{nharms}
\end{eqnarray}

 By employing the Pauli principle and the charge conjugation operation  one
 obtains~\cite{our-FB} that in the
$^1S_0$ channel the component $g_4$ is of the odd  parity
 while the remaining $g_1, ..., g_3$ are of the even parity.
 As mentioned, the BS amplitude is a mathematically
 complicated object and it is more convenient to considered it in Euclidean space, where
 the analytical properties of the amplitude become simpler and more transparent.
For convenience, in  Euclidean space we  redefine the odd partial
components $g_4$ for the $^1S_0$ channel as $ g_4\to ig_4$.
Then   the Wick rotated BS equation~(\ref{sphom})  reads
\begin{eqnarray}
g_n( p_4,|\bp|)= g^2  \int d \Omega _{p} \int\frac{d^4k}{(2\pi)^4}
S(k_4, |\bk|) \frac {1}{(p-k)^2+\mu^2}
\sum\limits_{m}A_{nm}(p,k)g_m(k_4,\bk) \label{par1s0},
\end{eqnarray}
where $m,n=1\ldots 4$ for the $^1S_0$    and the  scalar part
$S(k_4,|\bk|)$ of the two spinor propagators is defined as
\begin{eqnarray}
\label{prop} S(k_4,|\bk|)=\frac{1}{\left(k^2+  m^2-\frac{M^2}{4}
\right)^2+M^2k_4^2} \ .
\end{eqnarray}

   Next step is the expanding of the interaction kernel into hyperspherical
 harmonics~\cite{batman}

 \begin{eqnarray}
\frac {1}{(p-k)^2+\mu^2}&=&2\pi^2 \sum_{nlm} \frac{1}{n+1}
V_n(\tilde p,\tilde k)Z_{nlm}(\chi_p, \theta_p,\phi_p)
Z_{nlm}^{*}(\chi_k, \theta_k,\phi_k),\label{exp12}
\\
V_n(\tilde p,\tilde k)&=& \frac {4}{(\Lambda_+ +\Lambda_-)^2}
\left( \frac {\Lambda_+ -\Lambda_-}{\Lambda_+
+\Lambda_-}\right)^n;\quad
\Lambda_\pm = \sqrt{(\tilde p \pm \tilde k)^2+\mu^2}. \nonumber
\end{eqnarray}
with
 \begin{eqnarray}
Z_{nlm}(\chi,\theta,\phi)&=&X_{nl} (\chi) Y_{lm}(\theta,\phi);\,
X_{nl}(\chi)=\sqrt{\frac{2^{2l+1}}{\pi} \frac
{(n+1)(n-l)!l!^2}{(n+l+1)!}} \sin^l\chi C_{n-l}^{l+1}(\cos \chi),
\nonumber
\end{eqnarray}
where $Y_{lm}(\theta,\phi)$ are the familiar spherical harmonics,
and $C_{n-l}^{l+1}$ are the Gegenbauer polynomials.
The resulting system of equations reads
\begin{equation}
g_n( p_4,|\bp|)= g^2  \int\frac{k^3 dk \sin^2 \chi_k
d\chi_k}{(4\pi^3)} S(k_4, |\bk|) W_{l_n}(\tilde p,\tilde
k,\chi_p,\chi_k) \sum\limits_{m}a_{nm}(k_4,\bk)g_m(k_4,\bk)
\label{par2s0},
\end{equation}
where
\begin{eqnarray}
W_{l_n}(\tilde p,\tilde k,\chi_p,\chi_k) = \sum_l \frac{1}{l+1}
V_l(\tilde p,\tilde k) X_{l l_n}(\chi_p) X_{l l_n}(\chi_k).
\nonumber
\end{eqnarray}

Furthermore,   the partial vertex functions $g_n$ are expanded over the
basis $X_{nl}(\chi_p)$ as
\begin{eqnarray}&&
g_{1,2}(p_4,|\bp|)=\sum_{j=1}^\infty g_{1,2}^j(\tilde
p)\,X_{2j-2,0}(\chi_p)\label{s0p1};\quad
g_{3}(p_4,|\bp|)=\sum_{j=1}^\infty
g_{3}^j(\tilde p)\,X_{2j-1,1}(\chi_p);\\ &&
g_{4}(p_4,|\bp|)=\sum_{j=1}^\infty g_{4}^j(\tilde
p)\, X_{2j,1}(\chi_p). \label{s0p2}
\end{eqnarray}

  Placing eqs.~(\ref{s0p1}) and (\ref{s0p2}) in to eq.~(\ref{par2s0})
  and performing integration over hyper-angles analytically,
  the initial system of four-dimensional integration
  equations  reduces to a (infinite) system of only one-dimensional equations.
  Eventually, by a proper choice of  numerical integration method
  (the Gaussian formula, in our case)
  one easily transforms the latter system of integral equations in to
    an ordinary algebraic  system of
  linear  equations. Obviously, further solving procedure is straightforward
 (see for details Ref.~\cite{our-FB}).

\section {Results}
By the above procedure we solved the BS equation for a system of two spinors
with equal masses in $^1S_0$ state
interacting via exchanges of scalar and pseudoscalar mesons.
Herebelow, for the sake of brevity,  we present results only for scalar exchange mesons;
more detailed results can be found in~\cite{our-FB}. Note that the accuracy of the
proposed method depends up on the dimension $N_G$ of the Gaussian mesh used in
calculations and on number of terms $M_{max}$ used in the decomposition~(\ref{s0p1}) and (\ref{s0p2}).
Also the magnitude  of the coupling constants governs
the existence of the solution itself; at some critical values of
the coupling constants the solution of the BS equation may not exist
at all.

As an illustration of the stability of the numerical procedure, in
Table \ref{tab10}  we present results for the   masses of the bound state
$M(g^2)$  depending on the Gaussian mesh $N_G$
 and $M_\mathrm{max}$. Calculations have been performed for the
$^1S_0$ state,with a scalar meson exchange of  mass $\mu$ for two
values   $\mu=0.15\ \mathrm{GeV/c}^2$  and $\mu=0.5\
\mathrm{GeV/c}^2$; the constituent particles (nucleons) have been
taken with equal masses $m= 1.0\ \mathrm{GeV/c}^2$ for simplicity.
 Results presented in Table~\ref{tab10} clearly demonstrate
 that the approximate solution
 converges rather rapidly, and already
at $M_\mathrm{max}\sim 4-5$ and $N_G=64$ the method provides  a
good solution of the system.

\begin{table}[h]
~\centering\begin{tabular}{|c||c|c|c||c|c|c|} \hline $g^2=15$ &
\multicolumn{3}{c||}{$\mu=0.15~{GeV/c^2}$}
 & \multicolumn{3}{c|}{$\mu=0.5~ {GeV/c^2}$} \\
\hline $M_\mathrm{max}$  &$ N_G=32$ & $N_G=64$ & $N_G=96$
 & $N_G=32$ & $N_G=64$ & $N_G=96$ \\
\hline
\hline 1 & 1.9399 & 1.9399 & 1.9399 & 1.9984 & 1.9984 & 1.9984\\
\hline 2 & 1.9370 & 1.9370 & 1.9370 & 1.9982 & 1.9982 & 1.9982\\
\hline 3 & 1.9368 & 1.9368 & 1.9368 & 1.9982 & 1.9982 & 1.9982\\
\hline 4 & 1.9368 & 1.9368 & 1.9368 & 1.9982 & 1.9982 & 1.9982\\
\hline \multicolumn{7}{c}{}\\
\hline $g^2=30$ & \multicolumn{3}{c||}{$\mu=0.15~ {GeV/c^2}$}
 & \multicolumn{3}{c|}{$\mu=0.5~ {GeV/c^2}$} \\
\hline $M_\mathrm{max}$  &$ N_G=32$ & $N_G=64$ & $N_G=96$
& $N_G=32$ & $N_G=64$ & $N_G=96$ \\
\hline
\hline 1 & 1.7932 & 1.7910 & 1.7905 & 1.9167 & 1.9142 & 1.9137\\
\hline 2 & 1.7897 & 1.7875 & 1.7871 & 1.9152 & 1.9127 & 1.9122\\
\hline 3 & 1.7896 & 1.7874 & 1.7870 & 1.9152 & 1.9127 & 1.9122\\
\hline 4 & 1.7896 & 1.7874 & 1.7870 & 1.9152 & 1.9127 & 1.9122\\
\hline
\end{tabular}
\caption{Dependence of the bound state masses on the $M_{max}$ and on the Gaussian
mesh used in actual numerical calculations.}
\label{tab10}
\end{table}
The  solution of the BS equation, eq.~(\ref{spex}),
can be obtained  from the known numerical values of
 the  partial components $g_i$. In Fig.~\ref{pic1}
\begin{figure}[h]
\includegraphics[height=8cm,width=7cm]{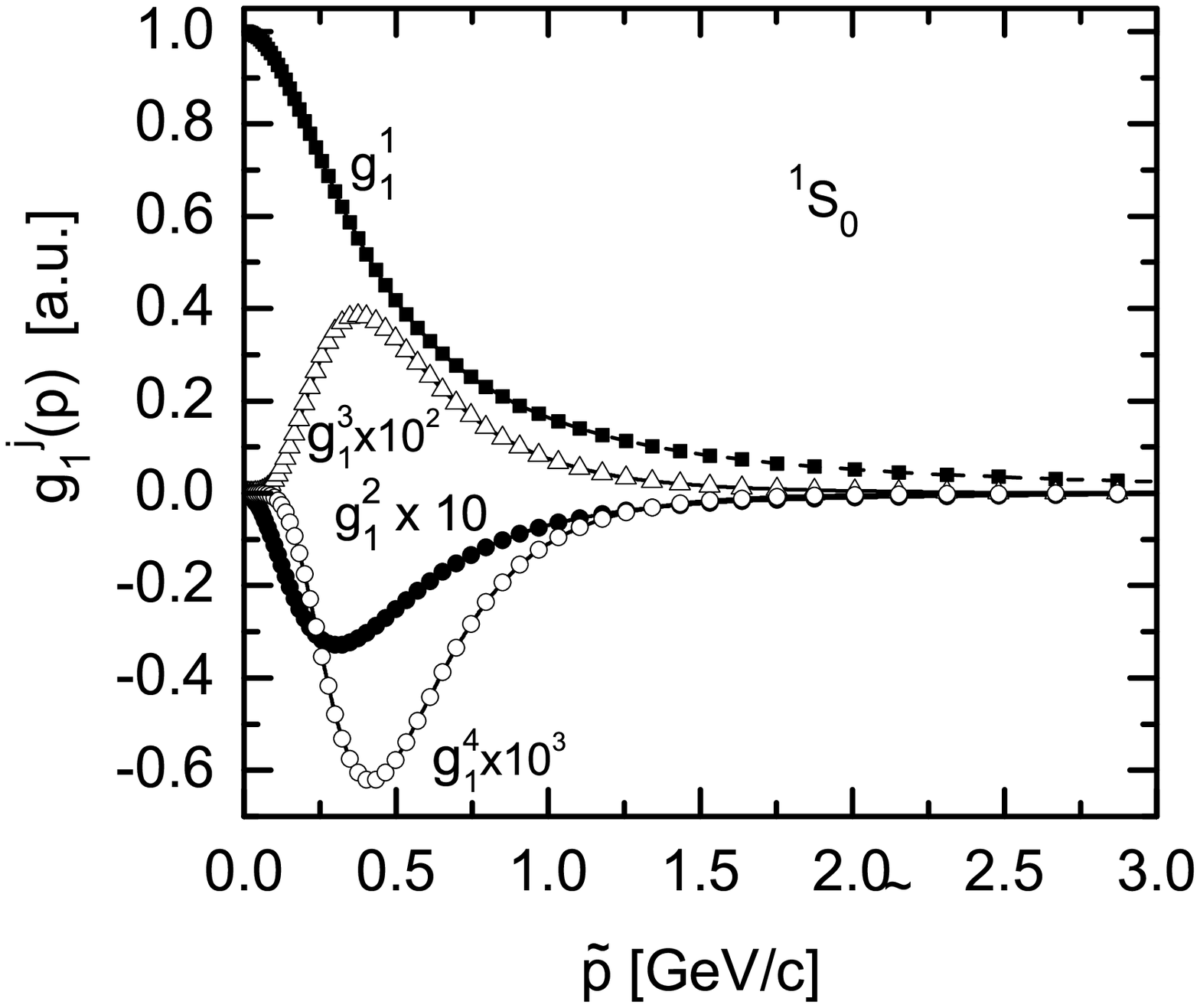}
\includegraphics[height=8.cm,width=7cm]{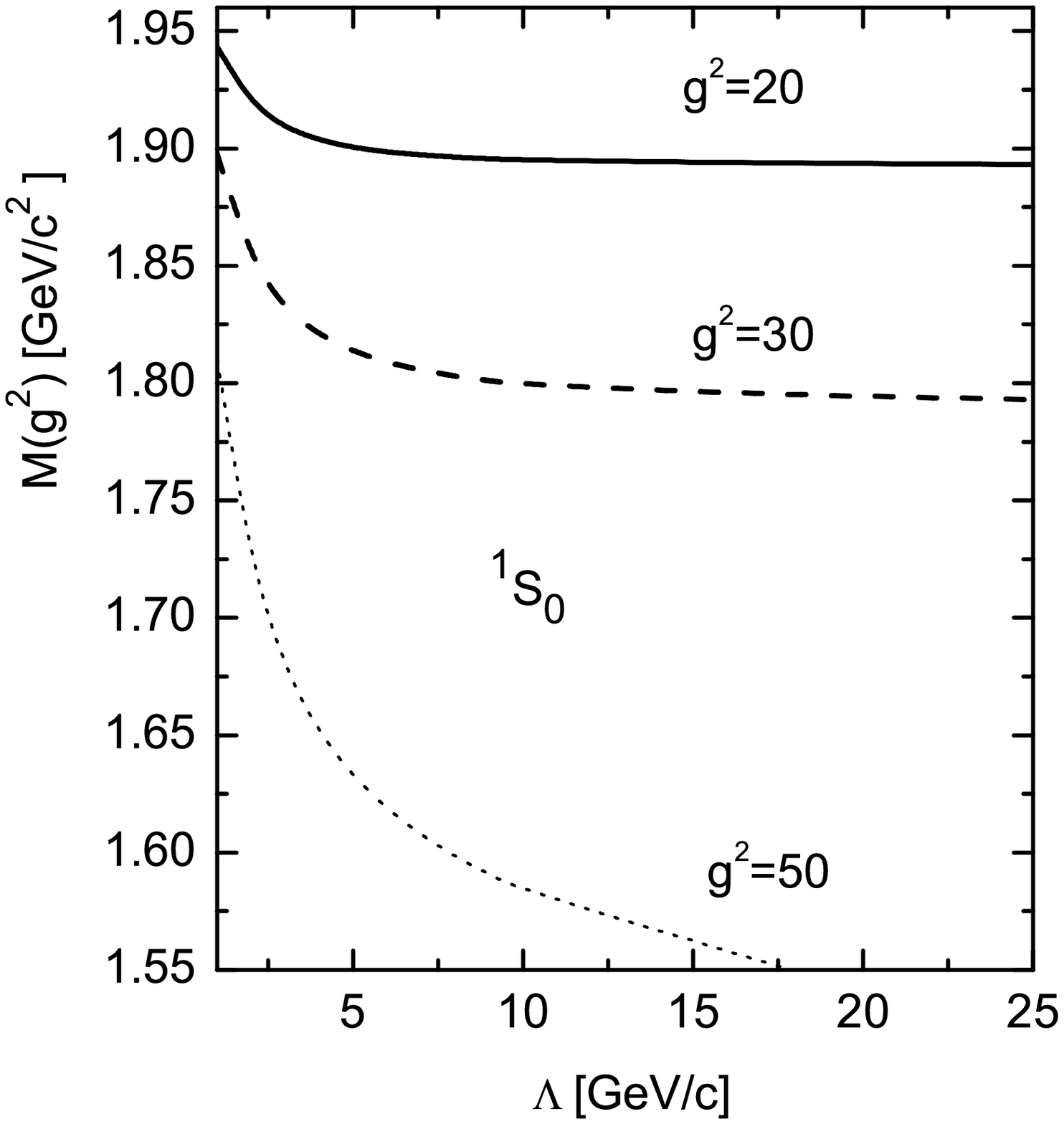}
\caption{{\it
Left panel: The coefficients
$g_1^j, j=1\ldots 4$,   eq.~(\ref{spex}),  as  functions of the
 euclidian momentum $\tilde p^2=p_4^2+{\bf p}^2$.
Right panel: The mass of the bound state $^1S_0$ as a function
of the cut-off parameter $\Lambda$ at different values of the coupling constant
$g$. At  large $g>g_{cr}$ the solution is strongly dependent on
the cut-off parameter, i.e. becomes rather unstable.}}
\label{pic1}
 \end{figure}
  we present the behavior of the partial coefficients  in the $^1S_0$ channel
  $g_1^j(\tilde p)$, $j=1 \ldots 4$, eq.  (\ref{s0p1}), as a function of the
  euclidian   relative momentum $\tilde p=\sqrt{p_4^2+{\bf p}^2}$.
  Calculations have been performed for a bound system with
   $M=1.937\ \mathrm{GeV/c}^2$. The actual parameters used in numerical calculations
  are  $N_G=96$, $M_\mathrm{max}=4$ and $g^2=15$.
 Closed squares correspond to
$g_1^1$,  closed circles - $g_1^2$ multiplied by 10, triangles
-$g_1^3$ multiplied by 100, open circles -$g_1^4$ multiplied by
1000.
 It can be seen that at large $\tilde p$  each function decreases
as inverse powers of $\tilde p$, which allows for a relatively simple parametrization
of the result;  the solid lines correspond to such a  fit.
An analysis of the obtained results shows that the numerical solution is rather sensitive
to the magnitude of the coupling constants. Moreover, it has been found that there are
some critical values of the coupling constants, $g^2_\mathrm{cr}$,
 above which the solution of the BS equation
does not exist (in absence of the cut-off form factors).
At values of the coupling constants close the their critical values
$g^2\sim g^2_\mathrm{cr}$
 the numerical solution becomes unstable and strongly dependent on the
 Gaussian mesh, $M_{max}$ and $\Lambda$. Such a situation is illustrated in
 \begin{wrapfigure}[15]{l}[0pt]{6cm}
\vspace*{-1cm}
\includegraphics[height=7cm,width=6cm]{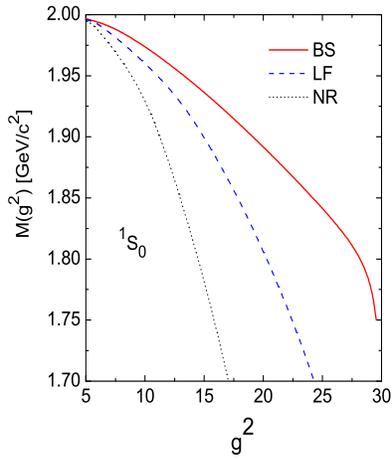}
\caption{ {\it The mass of the bound state $M$ in the $^1S_0$ channel
within different models,
non relativistic (NR), light front (LF) dynamics and present (BS) approach.}  } \label{pic5}
\end{wrapfigure}
 Fig.~\ref{pic1}, right panel, where we present the dependence of the mass $M$  on the
values of the cut-off parameter $\Lambda$ at different  coupling
constants $g^2$. It is clearly seen that at relatively low values of $g^2$ below its
critical value ($g^2_\mathrm{cr}\sim 40$)
the solution is practically independent on $\Lambda$, i.e. it converges rather rapidly.
 Contrarily,  at  $g^2 > g^2_\mathrm{cr}$ the
 solution $M(g^2)$ becomes manifestly   dependent on
  $\Lambda$, i.e., it can disappear at all. Such a behavior of the  solution
at $g^2\sim g^2_\mathrm{cr}$
 exactly reproduces the peculiarities of the well-known
 collapse phenomenon for  potentials like $-\alpha/r^2$ in
 nonrelativistic quantum mechanics.
 The magnitude of the critical value  $g^2_\mathrm{cr}$ may be estimated
 analytically and it turns out to be  $g^2_\mathrm{cr} \sim 4\pi^2$ (see Ref~\cite {our-FB}).
As mentioned, the obtained solution has a rather simple dependence on the
euclidian momentum $\tilde p$ and allows for a simple analytical parametrization
of the solution, which can be exploited further to perform a  Wick rotation
back to Minkowsky space. In this context is is instructive to compare
our solution with ones obtained within other known approaches.
In  Fig.~\ref{pic5} a comparison of our solution (solid line)
 with the results of
non relativistic calculations (dotted line)
as well as with the ones  within the Light Front
dynamics (LF)~\cite{karmanov} (dashed line) is presented.
As expected, at low values of the coupling constant different approaches provide similar results.
As $g$ increases the difference becomes more  significant, reflecting
the role of  relativistic effects. The difference between LF and BS approaches is
due to different treatment of the vertex function, namely within the LF the vertex function consists on
three components, while within the BS formalism four partial components describe the solution.
Obviously, the role of the fourth component increases with~$g^2$.

\vskip -2cm

\section { Conclusion}

We generalize a method based on hyperspherical harmonics to solve
the homogeneous spinor-spinor  Bethe-Salpeter equation in
Euclidean space. To do so, we introduce a new basis of
spin-angular harmonics, suitable to expand the Bethe-Salpeter
vertex into four-dimensional hyperspherical harmonics. We obtain
an explicit form of the corresponding system of one-dimensional
integral equations for the partial components and formulate a
proper  numerical algorithm to solve this system of equations. The
BS vertex functions are  studied in detail for
  the $^1S_0$ and $^3S_1-^3D_1$ bound states with  scalar, pseudoscalar
  and vector meson exchanges.
Our results are in a good agreement  with calculations within the
non relativistic and Light Front Dynamics approaches.
Within the novel method the effectiveness of the numerical
procedure is analyzed for the scalar, pseudoscalar and vector
meson exchanges and conditions for stability of the solution  are
established.

An advantage of the method is the possibility to present the
numerical solution in a reliable and simple analytical
parameterized form,  extremely convenient in practical
calculations of matrix elements within the BS formalism and for
analytical continuation of the solution back to Minkowski space.
The   method allows for a  covariant description of
two-body systems, such as the deuteron, positronium and the
variety of known mesons, as relativistic bound states
and, by solving  the inhomogeneous BS equation, to describe
the scattering states.

\end{document}